\documentclass[prl, aps,
twocolumn,
longbibliography,
amsmath,amssymb,
floatfix,
superscriptaddress
]{revtex4-2}

\usepackage{physics}
\usepackage{epsf}
\usepackage{float}
\usepackage{graphicx}
\usepackage{amsmath}
\usepackage[usenames]{color}
\usepackage{float}
\usepackage{bm}
\usepackage{relsize}
\usepackage[normalem]{ulem}

\def \be {\begin{equation}}
\def \ee {\end{equation}}
\def \bea {\begin{align}}
\def \eea {\end{align}}

\def \BEA {\begin{eqnarray}}
\def \EEA {\end{eqnarray}}
\def \BC {\begin{cases}}
\def \EC {\end{cases}}

\usepackage{siunitx}
\usepackage{tabularx}
\usepackage{xcolor}
\usepackage[version=4]{mhchem}
\usepackage{ctable}

\usepackage[unicode=true,colorlinks=true,citecolor=blue,urlcolor=blue]{hyperref}

\def\be {\begin{equation}}
\def\ee {\end{equation}}
\def\bea {\begin{align}}
\def\eea {\end{align}}

\def\bee{\begin{eqnarray}}
\def\eee{\end{eqnarray}}
\def\BC {\begin{cases}}
\def\EC {\end{cases}}

\def\Ediss {$E_\text{diss}$~}

\def\Edisso {$E_\text{diss}$.~}
\def\Df {$\Delta f$~}
\def\Dfx {$\Delta f(x)$~}

\def\Dfo {$\Delta f$.~}

\makeatother

\begin{document}
\title{
Atomic Identification via the Prandtl-Tomlinson 
Stick-Slip Mechanism
}

\author{S.~E.~Schweiss}
\affiliation{University of Regensburg, 93040 Regensburg, Germany}
\author{L.~Hoermann}\affiliation{University of Vienna, 1090 Vienna, Austria}
\author{F.~J.~Giessibl}
\affiliation{University of Regensburg, 93040 Regensburg, Germany}
\author{A.~J.~Weymouth}
\email{jay.weymouth@ur.de}
\affiliation{University of Regensburg, 93040 Regensburg, Germany}

\date{\today}

\begin{abstract}	

            Distinguishing atomic species is an ongoing challenge in the field of scanning probe microscopy. 
            Typically, atomic force microscopy makes use of electrostatic interactions to perform
            atomic identification. 
            Here we perform lateral force microscopy (LFM)  measurements with a CO terminated tip of a partially oxidized Cu(110) surface and show
            clear differences between the metal and oxygen sites over a large $z$-range.
            This allows for atomic identification on a typical metal oxide. 
            Density functional theory based calculations show that the tip-sample interaction is dominated by short-ranged orbital interactions and that the electrostatic contribution is negligible. 
            At small tip-sample distances, LFM can probe the Prandtl-Tomlinson stick-slip mechanism on the single-atom scale. 
            It is visible as a striking asymmetry in the forward and backward scan directions, experimentally revealing this fundamental model for nanotribology on the single-atom scale. 
\end{abstract}

\maketitle




Determining the atomic structure of surfaces is an active field of research with applications in catalysis and tribology~\cite{Zhou2020, RafsanjaniAbbasi2024, Zhou2022, DienwiebelPRL2004, Vazirisereshk2020}. 
Frequency-modulation atomic force microscopy (AFM) probes this 
by measuring tip-sample interactions with atomic resolution. 
State-of-the-art AFM relies on atomically well-defined tip terminations~\cite{Moenig2018, Mohn2013}, most commonly a single CO molecule (CO tip)~\cite{Gross2009}. 
Recently, frequency-modulation lateral force microscopy (LFM) combined lateral motion with the single-atom sensitivity of AFM allowing for single chemical bonds to be imaged~\cite{Giessibl2002, Weymouth2020, MartinJimenez2022}. 
Here, we demonstrate the use of LFM with a CO tip to precisely probe a potential energy surface (PES) and allow for atomic identification beyond that of a normal-force AFM with a CO tip~\cite{SchulzeLammers2021}.

The high sensitivity of normal-force AFM to the PES 
is observed in the decay length of the measurements which is on the order of a few tens of picometers~\cite{Ternes2011}. 
Dissipation measurements with CO tip LFM have a dramatically shorter decay length of 4\,pm~\cite{Weymouth2020}. 
During these dissipation measurements the CO tip, acting as a torsional spring, is deflected as it slides over a potential energy barrier until it snaps 
to the other side~\cite{Weymouth2020}.
This is effectively a realization of a one-dimensional Prandtl-Tomlinson model, central in understanding atomic-scale friction in nanotribology~\cite{Prandtl1928, Tomlinson01061929, Holscher2008}. 
In our setup, the O-atom at the end of the CO can be trapped in local potential energy minima of the surface as the metal apex of the tip moves laterally above the surface.
For oscillations including a snap in both sliding directions, work is done on the tip, providing direct experimental access to individual atomic-scale stick-slip events~\cite{Weymouth2020}. 


In this Letter, we conduct LFM on the Cu(110)-(2$\times$1)O added-row (AR) reconstruction, a prototypical one-dimensional corrugated potential. 
We demonstrate that atomic identification on an oxide surface can be achieved with a CO tip. 
A model based on Lennard-Jones interactions (LJ-simulation) allows the CO tip's sliding motion to be analysed and the spatial extent of the measured dissipation signal to be understood.
Density-functional-theory (DFT) calculations reveal that the interaction is governed primarily by direct orbital interactions resulting in atomic contrast.  
By approaching sufficiently close to the surface, we further observe an asymmetry in the frequency-shift ($\Delta f$) signal between forward and backward scan directions, providing direct experimental evidence for the predicted stick-slip motion of the tip apex over an atomic-scale potential barrier. 
These results establish LFM as a powerful experimental platform for investigating the elementary stick-slip processes underlying atomic-scale friction.

Experimental data were acquired using a combined STM/AFM system (CreaTec Fischer \& Co. GmbH). To perform STM and LFM measurements, a qPlus sensor \cite{giessibl1998} was mounted such that it oscillates laterally to the surface~\cite{Giessibl2002}. 
All measurements were performed at $5.3$~K and in UHV conditions. Standard sputter and anneal cycles were used to prepare the bare Cu(110) surface. 
The sample was then partially oxidized by keeping it at a temperature of $250~{^\circ}$C while leaking in molecular oxygen at a pressure of $1\times10^{-8}$~mbar for less than 1\,s. 
Subsequently, the sample was annealed at $450~{^\circ}$C for 20 minutes~\cite{Monig2018, yesilpinar2021mechanical}. 
After cooling the sample, CO molecules were introduced and used to functionalize the tip apex~\cite{Gross2009, Bartels1997}. 
The quality of the CO tip was then assessed using the CO front atom identification (COFI) method~\cite{Emmrich2015, Welker2012, Gretz2020}. 
All LFM images shown are constant-height and measured with an applied bias of $1$~mV and an amplitude of $50$~pm. 
More details are found in the Supplemental Material.

\begin{figure}
	\centering \includegraphics[width=1\linewidth]{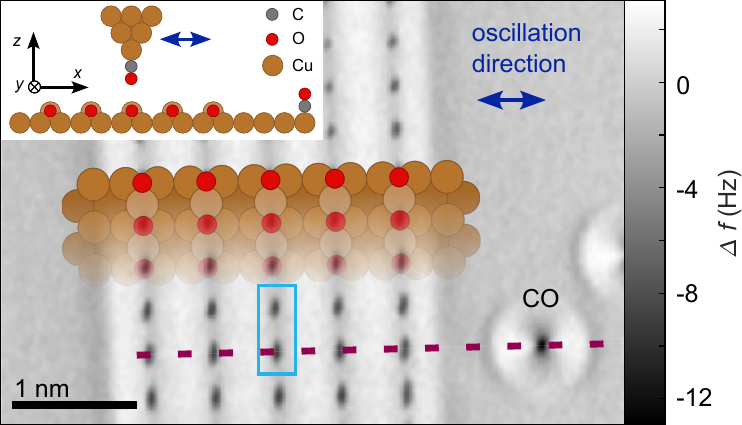}
	\caption{\Df image of partially oxidized Cu(110)-($2\times 1$)O with the added row (AR) reconstruction. 
    A Cu oxide stripe is visible as well as two CO molecules. 
    The blue double arrow indicates the oscillation direction of the tip, the purple dashed line guides the eye for feature alignment. A schematic of the surface is overlaid, indicating the position of the atomic species. For orientation purposes, the light blue box indicates a frame of the middle AR. The inset shows a side view of the surface schematically together with a CO terminated tip (Cu-apex ending in a CO) and a CO on the bare Cu(110).}
	\label{Fig1} 
\end{figure}

Figure \ref{Fig1} shows a constant-height \Df overview scan of the partially oxidized Cu(110) surface 
measured with a CO tip. 
An oxide stripe consisting of five ARs can be seen, as well as two CO molecules on the bare Cu surface. The oscillation direction is indicated by a blue double arrow. The ARs can be seen as distinct dark striped lines. 
The nearby surface CO molecule can be used to identify the atomic species corresponding to these dark features. 
CO molecules adsorb on top sites of the bare Cu(110) surface~\cite{okabayashi2023dynamic}, 
which are aligned with the AR-O sites. 
The purple dashed line in Fig.~\ref{Fig1} lies above the top sites, guiding the eye and identifying the AR-O within the stripes. 

\begin{figure}
	\centering \includegraphics[width=1\linewidth]{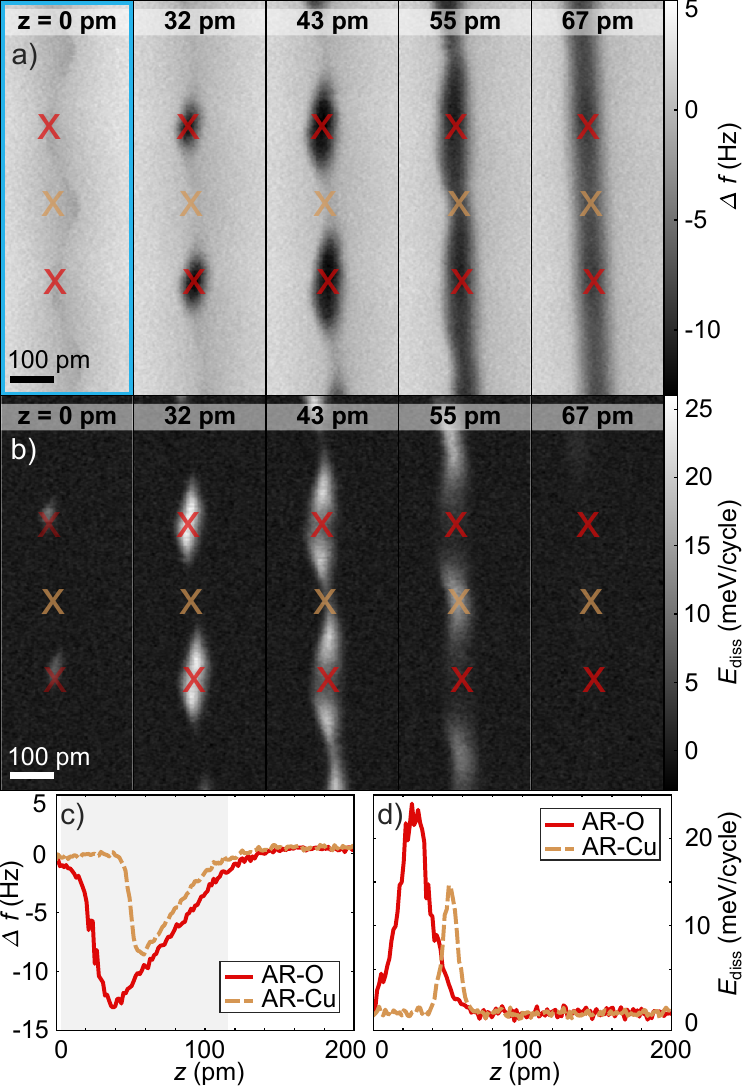}
	\caption{a) \Df and b) \Ediss images of one AR for multiple $z$, with $z = 0$~pm denoting the closest measurement. The light blue box corresponds to an equivalent frame as the light blue box in Figure \ref{Fig1}. The atomic species are indicated by red (O) and brown (Cu) crosses. $z$-spectra of c) \Df and d) \Ediss over AR-O and AR-Cu sites. The gray area in c) denotes the range over which the two atomic species can be clearly distinguished.
    }
	\label{Fig2} 
\end{figure}

In Figure \ref{Fig2} a) and b), the $z$-dependence of an AR is depicted in both \Df and energy dissipation ($E_\text{diss}$). 
Here, $z = 0$~pm denotes the closest measurement. 
The \Df images in \ref{Fig2} a) show the dark features discussed in Figure \ref{Fig1} appearing over AR-O sites for $z$ around $20-50$~pm. 
Further slices are shown in the Supplemental Material Fig.~S1.
For $z>50$~pm, these features start to broaden in $y$-direction while keeping the minimum at the AR-O position. 
For $z\approx 55$~pm, the dark lines appear similar to chemical bonds that were previously reported~\cite{Weymouth2020, MartinJimenez2022}.
Dark features in \Df indicate a maximum in potential energy~\cite{Weymouth2018}. 
The simultaneously measured \Ediss channel can be seen in \ref{Fig2} b). 
For $z<40$~pm, maxima occur over the AR-O positions, while for $z>50$~pm they occur over the AR-Cu positions. 
For intermediate $z$, the \Ediss features appear in between the AR-O and AR-Cu sites. 

The position of the local \Ediss maximum thus changes as a function of $z$.
This is in contrast to observations of \Ediss over flat chemical bonds, where the \Ediss maximum at each height is centered 
between the two atoms of that bond~\cite{Weymouth2020}. 

Figure~\ref{Fig2} c) and d) are $z$-spectra of \Df and \Ediss over the AR-O and AR-Cu sites.
 These plots demonstrate a clear difference between the atomic species of the AR in both \Df and \Edisso 
 The \Df curves show that AR-O can be clearly distinguished from AR-Cu over a large $z$-range of approximately $115$~pm, indicated by the gray area in Figure \ref{Fig2} c). 
 The AR-O curve shows a broader and deeper minimum than the AR-Cu curve. It lies below the AR-Cu curve for the full range of interest. 

This distinguishability is particularly surprising as atomic identification over metal oxides with a CO tip using normal-force AFM has been challenging~\cite{SchulzeLammers2021}.
It has been proposed that clear identification is only possible by using tip apices with much stronger electrostatic fields, such as CuOx tips~\cite{Hutner2024}.
Our results suggest that CO tips suffice to clearly observe differences between the metal and oxygen in a metal oxide. 

\begin{figure}
	\centering \includegraphics[width=1\linewidth]{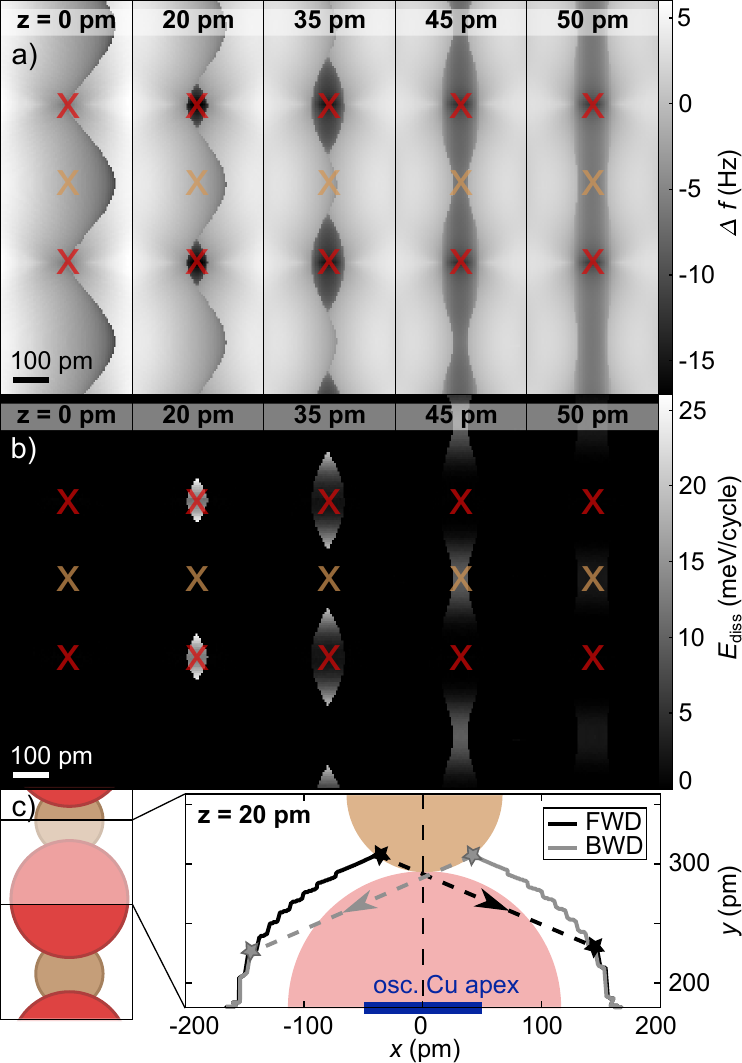}
	\caption{Simulated a) \Df and b) \Ediss images over one AR for multiple $z$. $z = 0$~pm denotes the height corresponding to the closest measurement, other $z$-values are relative to that. The atomic species are indicated by red (O) and brown (Cu) crosses. c) Schematic zoom in to the AR. The trace of the tip-O atom is plotted during the forward (black) and backward (gray) oscillation of the Cu-apex. The Cu-apex is oscillating in the middle of an AR-O site (blue). Snapping from one side of the AR to the other is indicated by stars.}
	\label{Fig3} 
\end{figure}

To investigate the position change of the \Ediss maximum as a function of $z$, we simulated \Df and \Ediss using 
a LJ-simulation~\cite{NamPNAS2024}. 
Details of the simulation can be found in the Supplemental Material.

Figure \ref{Fig3} a) and b) show the simulated \Df and \Edisso 
The tip heights are chosen to qualitatively reproduce the experimental data shown in \ref{Fig2} a) and b). 

Very good agreement of both \Df and \Ediss indicates that the simulation accurately describes the location of the tip-O atom during the oscillation. 
Figure \ref{Fig3} c) shows the position of the apex-O atom 
during one oscillation of the Cu-apex over the AR-O for small $z$. 
The apex-O atom bends around the AR-O until it snaps to the other side of the AR, indicated by stars. 
This snapping motion is the root cause of the \Edisso 
More snapping scenarios can be found in the Supplemental Material Figure S2. 

\begin{figure*}
	\centering \includegraphics[width=1\linewidth]{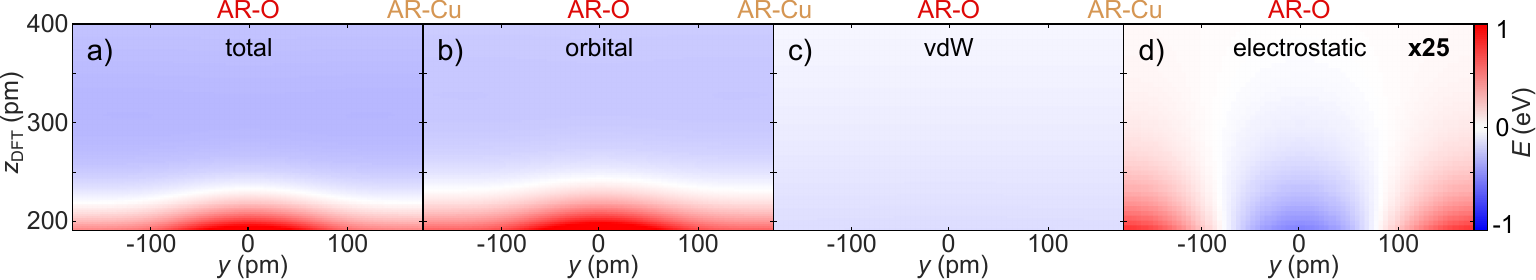}
	\caption{Calculated potential energy surface (PES) over the AR. a) Combined PES consisting of b) short-range orbital contribution and c) long-range vdW contribution. d) Electrostatic contribution via Hartree potential of $0.05$ electron charge, multiplied by a factor of  $25$. 
    }
	\label{Fig4} 
\end{figure*}

To understand the fundamental interaction between the tip and surface, we performed DFT simulations. 
Details can be found in the Supplemental Material. 
In Figure \ref{Fig4} a) the calculated PES over the $y$-direction of one AR is shown. A vertically-oriented CO molecule acts as the tip. It is attached to one Cu atom, imitating the bulk background of the tip. 
The PES shows that the AR-O site is more repulsive than the AR-Cu site at close distances, in agreement with the chosen LJ-parameters.

Figure \ref{Fig4} b) shows the contribution from the short-range interaction to the PES, as modelled by the exchange-correlation functional.

Figure \ref{Fig4} c) shows the long-range dispersion contribution, which can be interpreted as the van der Waals (vdW) interaction~\cite{hormann2020reproducibility}.

Figure \ref{Fig4} d) shows the electrostatic contribution. The DFT-calculated electrostatic voltage above the surface was multiplied by  0.05 elementary charge, simulating the charge at the apex of the CO tip~\cite{VanderLit2016}. The contrast must be multiplied by a factor of 25 to see any features with this color scale.

Therefore, the corrugation of the PES is a result of the dominant short-range interaction, while the vdW interaction contributes a weak attractive background.
The electrostatic contribution is negligible.

Previous studies have made use of the electrostatic interaction to perform atomic identification.
For instance, Schulz et al. identified the B and N sites of BN with a CO tip using normal-force AFM~\cite{Schulz2018a}.
There, the large dipole moment of the metal tip contributed to the atomic contrast.
Relying on electrostatics for atomic identification is why the CuOx tip, with its larger apex charge, has been successful in performing atomic identification of metal oxide surfaces~\cite{Wiesener2024, Hutner2024}. 

Instead of the electrostatic interaction, 
we are most sensitive to short-range interactions as established by Moll et al. to explain the high spatial resolution achieved with CO tips~\cite{Moll2010, Moll2012}.
Here we show that although LFM and AFM with a CO tip both probe the same PES over the AR, LFM is able to provide clear atomic identification where normal-force AFM is not~\cite{SchulzeLammers2021}. 


To explain the enhanced sensitivity of LFM we consider that the negative \Df signal over the AR indicates 
a potential energy maximum.
On one side of the AR, the lateral forces are oriented in one direction and on the other side, in the other direction, maximizing the observed \Dfo 
In normal-force AFM, however, the tip apex slides up and down on one side of the potential energy barrier.
Further discussion is in the Supplemental Material Figure S4.


\begin{figure}
	\centering \includegraphics[width=1\linewidth]{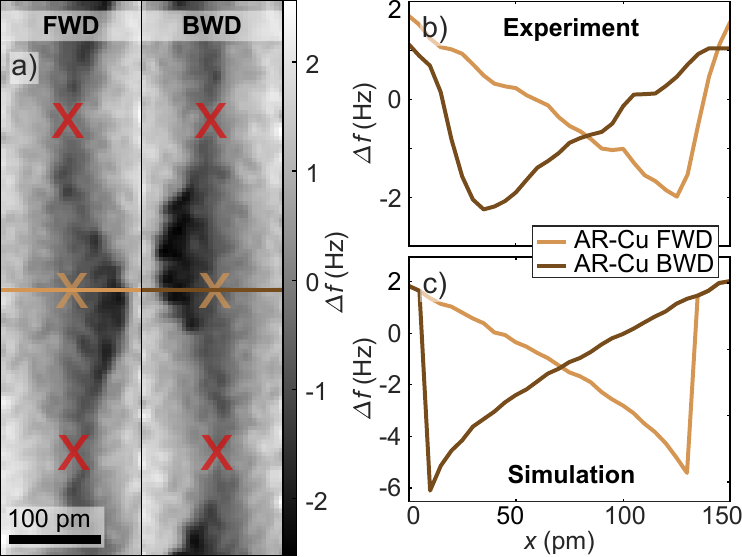}
	\caption{a) Experimental \Df images (forward and backward) for $z = 0$~pm. The atomic positions are indicated by red (O) and brown (Cu) crosses. b) Experimental and c) simulated \Dfx forward and backward line scans at AR-Cu site. The locations of the \Dfx slices are indicated by brown lines in a).}
	\label{Fig5} 
\end{figure}

\Df images taken at $z<40\,$pm show a striking left-right asymmetry in both experiment and simulation, as shown in Figures \ref{Fig2} and \ref{Fig3} a). 
At $z = 0$~pm little to no signal is observed in \Edisso 
Plotting the forward and backward \Df scan directions individually, see Figure \ref{Fig5} a), shows a mirror symmetry that is most prominent at AR-Cu sites. 
Plotting the \Dfx line scans that run above the AR-Cu, see Figure \ref{Fig5} b), reveals a hysteresis between the forward and backward scans. 
The simulated \Dfx curves, see Figure \ref{Fig5} c), also show this hysteresis. The Supplemental Material provides line scans at further heights for the AR-Cu and AR-O sites in experiment and simulation, see Figures S5 and S6. 

At these locations where forward-backward asymmetry is observed, the CO molecule is trapped on one side of the AR.
Here, we do not expect an \Ediss signal because we are not closing a force-distance hysteresis loop with every oscillation cycle.
Instead, the \Df signal records this trapping of the CO tip as an asymmetry. 
This observation is evidence of the proposed Prandtl-Tomlinson stick-slip mechanism~\cite{Weymouth2020, Nam2026}.

In summary, we report upon LFM measurements 
with a CO tip over a partially oxidized Cu(110) surface.
We clearly show that the metal and oxygen sites of the oxide row can be resolved and distinguished. 
The \Df measurements show differences over a large $z$-range of $\approx 115$~pm. 
\Ediss measurements can also be used to distinguish the metal and oxygen sites. 
A LJ-simulation convincingly recreates the experimental output allowing access to the CO dynamics.
DFT based simulations confirm that the tip-sample interaction is dominated by the direct orbital interaction and not electrostatics. 

Typically AFM studies make use of electrostatics to perform atomic identification~\cite{Kyeyune2024, Hutner2024, Wiesener2024}, but we show that LFM can identify atomic species based instead on short-range orbital interaction.
This opens the possibility for atomic identification on surfaces which contain a large number of different atomic species with similar local charges.

At small $z$, the \Df images exhibit a pronounced asymmetry between the forward and backward scan direction over the AR-Cu site. 
This is a direct measure of atomic-scale stick-slip motion as described by the Prandtl-Tomlinson model, which itself is used by many nanotribology studies to
interpret their findings 
~\cite{Vazirisereshk2020, Li2024, Ying2025}.
Experimentally, many of these studies use friction force microscopy~\cite{MatePRL1987}, 
where the measured 
force arises from 
many-atom tip-surface interactions. 
%
LFM with an engineered single-atom tip apex can realize the idealized Prandtl-Tomlinson model, which is a cornerstone of nanotribology~\cite{Gnecco2012, Holscher2008, Dong2011}.

Financial support is gratefully acknowledged from the German Research Foundation (DFG, Project number 529680507).



%

\end{document}


\title{Atomic Identification via the Prandtl-Tomlinson Stick-Slip Mechanism}

\author{S.~E.~Schweiss}
\affiliation{University of Regensburg, 93040 Regensburg, Germany}
\author{L.~Hoermann}
\affiliation{University of Vienna, 1090 Vienna, Austria}
\author{F.~J.~Giessibl}
\affiliation{University of Regensburg, 93040 Regensburg, Germany}

\author{A.~J.~Weymouth}
\email{jay.weymouth@ur.de}
\affiliation{University of Regensburg, 93040 Regensburg, Germany}

\date{\today}

\maketitle
\tableofcontents
\clearpage
\section{Experimental Details}

The sample, a Cu(110) single crystal, is cleaned in UHV by repeated cycles of $\text{Ar}^+$ sputtering and annealing at $450~^\circ$C, both steps lasting for $20$~min.
Following the last cycle, the sample is partially oxidized by keeping a temperature of $250~^\circ$C while dosing molecular $\text{O}_2$ to a pressure of $P\approx 1\times10^{-8}$~mbar for less than $1$~s. 
A last anneal is performed at, $450~^\circ$C for $20$~min. 
The sample is oriented such that the resulting oxide rows are approximately perpendicular to the oscillation direction of the sensor. 
CO molecules are introduced to the surface at low temperature, $T = 5.3$~K. 

All experiments are conducted using a specially mounted qPlus sensor that allows LFM measurements~\cite{Giessibl2002}. 
A chemically etched tungsten tip is attached to the sensor. 
It is further sharpened by performing a field emission in UHV and poking into the bare Cu(110) surface. The quality of the tip apex is assessed via the CO front atom identification (COFI) method, using CO molecules on the bare Cu(110) surface. After that the tip apex is functionalized with a CO molecule. On this partially oxidized Cu(110) surface, the CO pickup is done by performing large overview scans in STM feedback that include oxide stripes. CO spontaneously attaches to a monoatomic Cu tip while scanning. The COFI method is again used to check the tip apex for symmetry and stability. 

The sensor possesses a resonance frequency of $f_0 = 39831$~Hz, a stiffness of $k = 899$~N/m, and a quality factor of $Q = 62948$. 

All experimental images were lightly filtered using the MATLAB imgaussfilt, which executes a two-dimensional Gauss filter with a standard deviation of 1 pixel. For the data shown in Fig. 2 of the main text, each pixel has a size of 2.3~pm $\times$ 1.8~pm.

\subsection{Noise Calculation}

The main source of noise in our setup is detector noise, typical for qPlus measurements~\cite{hembacher2002evaluation}. It applies to the \Df data and can be described via the minimum detectable force gradient~\cite{wutscher2011atomic}:
\begin{align}
    \delta k_{det} = \sqrt{\frac{8}{3}} \frac{n_q k B^{3/2}}{A f_0},
\end{align}
with the electrical noise $n_q$, the spring constant of the sensor $k$, the detector bandwidth $B$, the oscillation amplitude $A$, and the resonance frequency $f_0$. 

With values of both sensor specific ($k = 899.1784$~$\text{fm}/\sqrt{\text{Hz}}$, $f_0 = 39831$~Hz) and operating parameters ($B = 50$~Hz, $A = 50$~pm), we get a noise of $\delta k_{det} \approx 0.0209$~N/m. The raw data therefore has an uncertainty of the frequency shift:
\begin{align}
    \delta\Delta f = \frac{f_0}{2 k} \delta k_{det} \approx 0.462~\text{Hz}.
\end{align}

The uncertainty in the data shown in the paper is smaller than this value given the light filtering of the data mentioned above.

\subsection{Additional Experimental Data}

\begin{figure}
	\centering \includegraphics[width=1\linewidth]{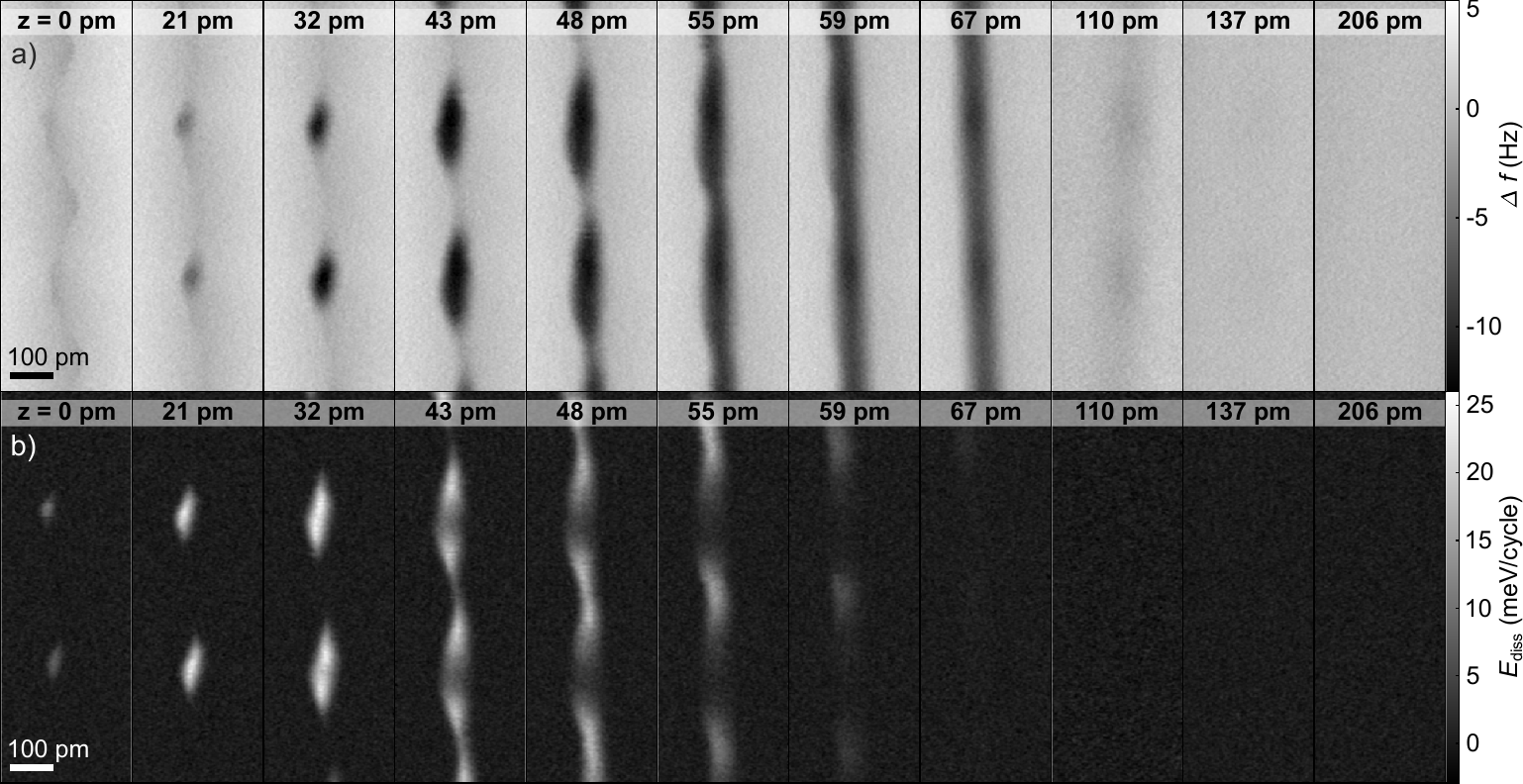}
	\caption{Constant height LFM images of one AR for multiple $z$. a) Shows the \Df and b) the \Ediss signal. The features for both \Df and \Ediss fade in and out within this $z$-range.}
	\label{FigS1} 
\end{figure}

Figure \ref{FigS1} shows additional constant height LFM images, i.e. a) \Df and b) \Ediss, for $z=(20.6,~48,~59,~109.8,~137.3,$ $205.9)$~pm. 
These images are of the same dataset as for Figure 2 a) and b) of the main text. 
In Figure \ref{FigS1} a) the additional images more clearly show the onset of the AR-O \Df features at small $z$ and their steady broadening until $z = 67$~pm. 
The fading of the broadened features can then be observed for $z = 110$~pm and beyond. 

The additional (compared to the main text) \Ediss images in Figure \ref{FigS1} b) show clearer the smooth progression of the features. 
The maximum of the \Ediss appears at the AR-O site for small $z$ and increases in intensity until $z = 31.6$~pm. 
After that, the features starts to split until pairs of local maxima merge above the AR-Cu sites at a height around $z = 59$~pm. 
Above this point, the intensity gradually fades, until no feature is visible for $z = 67$~pm and above. 

\clearpage

\section{Computational Details of the Lennard-Jones Simulation}

To replicate the experimental LFM results, a simulation based on the Lennard-Jones interaction between the tip and the surface is used (LJ-simulation).
We have used this model to simulate LFM results with a CO tip in previous publications~\cite{NamPNAS2024, Nam2026}. 
Similar to previous simulations for friction force microscopy, it includes a term that describes the interaction of the tip apex with the surface and the flexibility of the tip apex~\cite{Holscher1997}. 

The interaction of the surface is described by a sum of atomic interactions. 
They include interactions of each atom with the apex O atom of the CO at the tip apex. 
Similar to the first iteration of the probe-particle model~\cite{Hapala2014}, these are described by the Lennard-Jones empirical formula requiring the atomic radius and bond energy.

Excellent agreement for the experimental data were found with bond energies of 10\,meV. 
The radii of the Cu in the oxide and the O at the tip apex was set to 166\,pm. 
The radius of the O in the oxide was 223\,pm.

At each position of the metal tip apex, the CO is allowed to relax as a torsional spring under the interaction with the surface.
Then the total force on the apex is determined and from this, both \Df and \Edisso

Input parameters include the length of the moment arm of the CO molecule (300\,pm), the torsional stiffness of the tip ($25$~zNm), the resonance frequency $f_0$, the cantilever stiffness $k$, and the oscillation amplitude $A$.

\clearpage

\section{Additional Lennard-Jones Simulated Snapping Scenarios}

\begin{figure}
	\centering \includegraphics[width=1\linewidth]{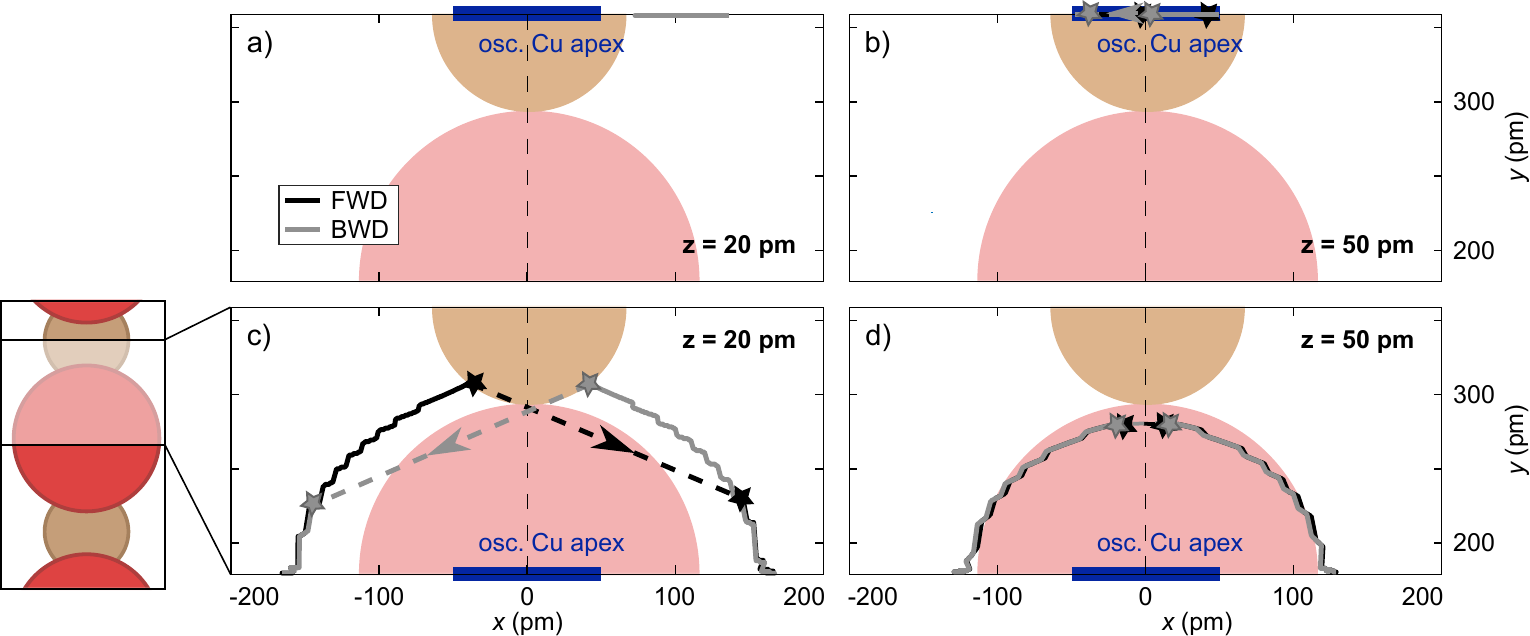}
	\caption{L-J simulated snapping dynamics of the apex-O atom over the AR. a) and c) show traces for an oscillation of the Cu-apex (blue) over the AR-Cu and AR-O sites, respectively, for $z = 20$~pm. b) and d) show the same for $z = 50$~pm. Snaps are indicated by stars. c) is identical to Figure 3 c) of the main text.}
	\label{FigS4} 
\end{figure}

To further investigate the L-J simulated CO-snapping paths introduced in the main text Figure 3 c), Figure \ref{FigS4} shows additional traces of the apex-O being oscillated over the AR at multiple locations and heights. 
In Figure \ref{FigS4} a), the Cu-apex oscillates over the AR-Cu site at a relative height of $z = 20$~pm. The resulting trace of the O-apex is confined to the right-hand side of the AR. 
There are no snaps present, so the lack of \Ediss signal observed in main text Figures 2 and 3 b) for the AR-Cu site at low $z$ can be understood. 
In Figure \ref{FigS4} b), the Cu-apex still oscillates over the AR-Cu site, but at a height of $z = 50$~pm. 
Here, the trace of the O-apex progresses over the center of the AR-Cu site, similar to the Cu-apex oscillation. 
Snaps are observed, explaining the presence of an \Ediss signal at the AR-Cu site for larger $z$, observed in main text Figures 2 and 3 b). 
Identical to main text Figure 3 c), Figure \ref{FigS4} c) depicts the O-apex dynamics for an oscillation of the Cu-apex over the AR-O site at $z = 20$~pm. 
The O-apex traces around the AR-O site, before it snaps over to the other side of the AR. 
The snap is possible once the apex-O reaches the AR-Cu site. 
The snap is rather large and as expected, a large signal in \Ediss is observed for this height and lateral position, see main text Figures 2 and 3 b). 
Finally, in Figure \ref{FigS4} d), the Cu-apex oscillates over the AR-O site for $z = 50$~pm. 
Like in the previous case, the O-apex traces around the AR-O site, but is not deflected as much. 
It is located closer to the center of the AR. 
Snapping then occurs close to the very middle of the AR. 
The snap is very small, explaining the barely noticeable \Ediss in the main text Figure 3 b) for this position. 

Overall, understanding the path of the O at the apex of the CO tip aids the interpretation of the \Ediss maximum with $z$ observed in experiment and simulation, see main text Figures 2 and 3 b). 
At varying $z$, the O-apex gets trapped or deflected differently, resulting in this characteristic $z$-progression.

\clearpage

\section{Computational Details of the DFT Simulation}

All electronic structure calculations were performed using the all-electron, numeric atom-centered orbital code FHI-aims~\cite{blum2009ab}. 
Exchange and correlation effects were treated using the HSE06 hybrid functional~\cite{krukau2006influence}, using the default mixing fraction of 0.11 and a screening parameter unit set to inverse Bohr. Long-range van der Waals interactions were accounted for via the non-local many-body dispersion (MBD-NL) correction~\cite{ambrosetti2014long}.

The systems were modeled with a net charge of zero and a closed-shell electronic configuration without spin polarization. Scalar relativistic effects were included via the atomic zeroth-order regular approximation (ZORA). Fractional occupancies at the Fermi level were treated using Gaussian smearing with a broadening width of 0.05~eV. 

The surface was modelled using the repeated slab approach, with a vacuum distance of $60$~\AA. The Brillouin zone of the primitive surface slab was sampled using a dense $52 \times 36 \times 1$ $\mathbf{k}$-point grid, which is equivalent to $64 \times 64 \times 64$ $\mathbf{k}$-grid for a Cu bulk unit cell. To account for macroscopic asymmetry and prevent spurious electrostatic interactions across the vacuum gap, a dipole correction was explicitly enabled, alongside multipole error compensation to ensure accurate electrostatic limits.

Geometry optimizations were performed using the trust radius method (TRM). Atomic positions were relaxed until the maximum residual forces on all unconstrained atoms fell below $10^{-2}$~eV/\AA. As for the basis set, the 2020 default tight basis functions were used.

\subsection{Convergence tests}

\begin{figure}[htpb]
    \centering
    \begin{tabular}{cc}
         \includegraphics[width=0.5\linewidth]{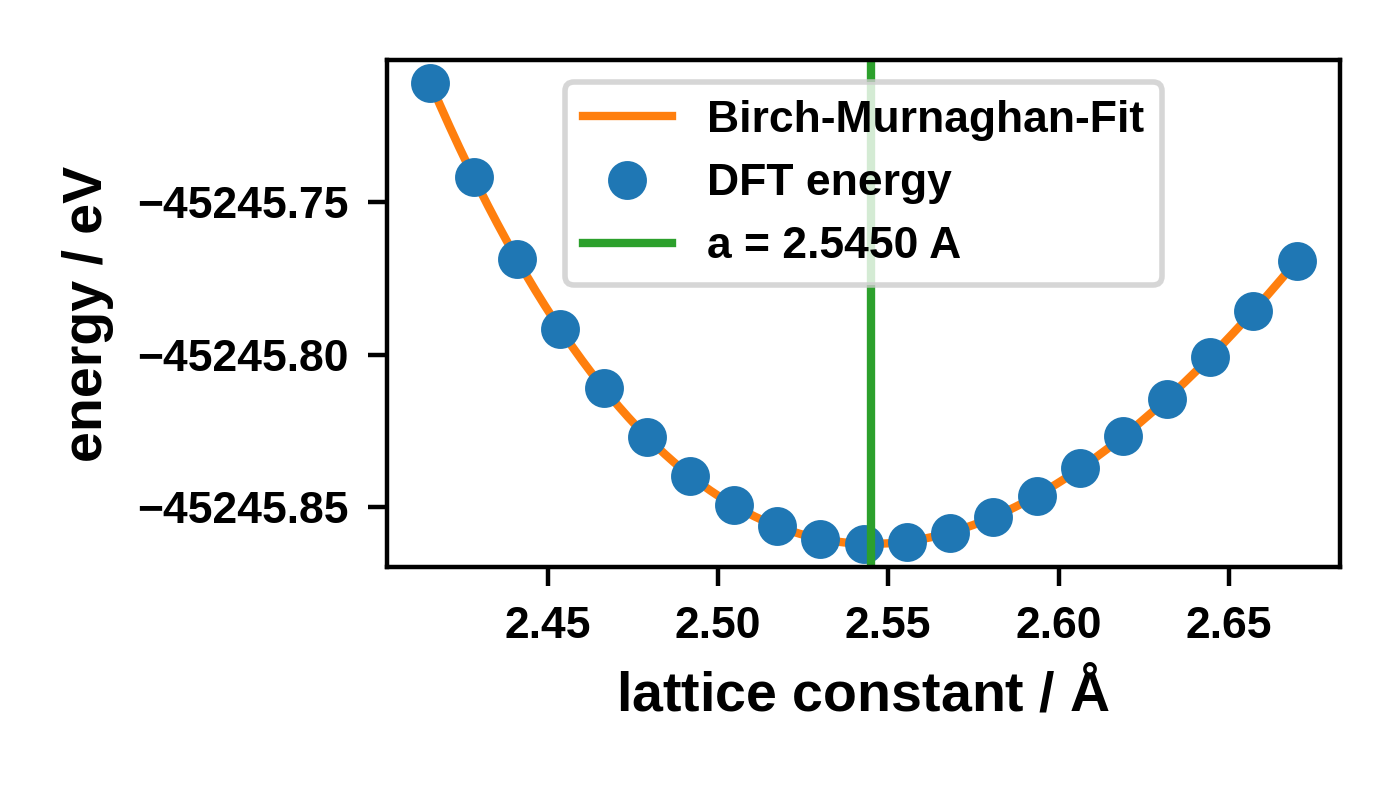} &
         \includegraphics[width=0.5\linewidth]{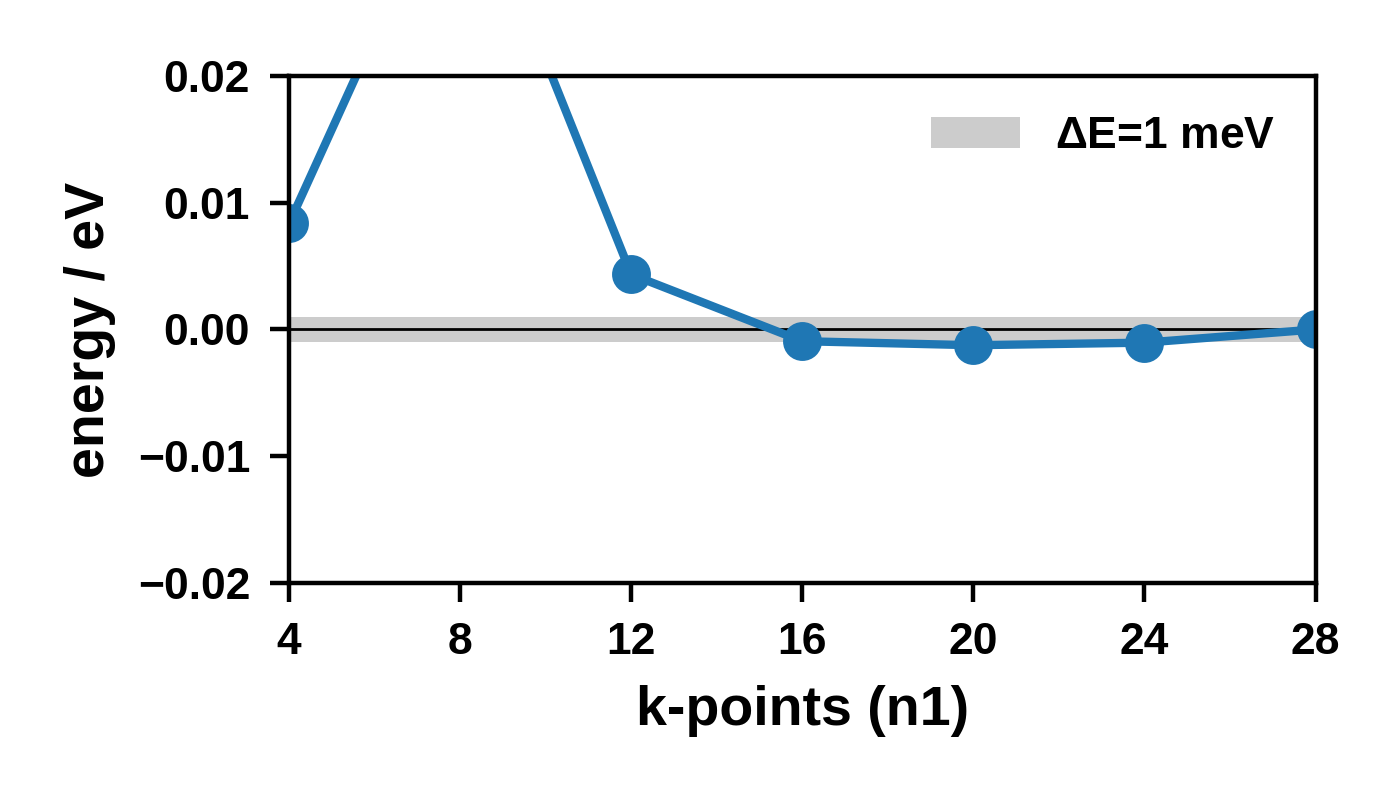}
    \end{tabular}
    
    \caption{Left: Total energy as a function of the lattice constant for the Cu primitive bulk unit cell; Right: Convergence of the total energy with respect to the $\mathbf{k}$-point grid density for the Cu primitive bulk unit cell.}
    \label{fig:convergence}
\end{figure}

Convergence tests were performed on the Cu primitive bulk unit cell using the computational settings described above. To determine the equilibrium structure, a dense $64 \times 64 \times 64$ $\mathbf{k}$-point grid was employed, yielding an equilibrium lattice constant of $2.545$~\AA\ (Figure~\ref{fig:convergence} Left). Furthermore, Brillouin zone sampling tests confirm that the total energy is fully converged at this $64 \times 64 \times 64$ $\mathbf{k}$-grid density (Figure~\ref{fig:convergence} Right).

\clearpage

\section{Comparison LFM and Normal-Force AFM}

\begin{figure}
	\centering \includegraphics[width=1\linewidth]{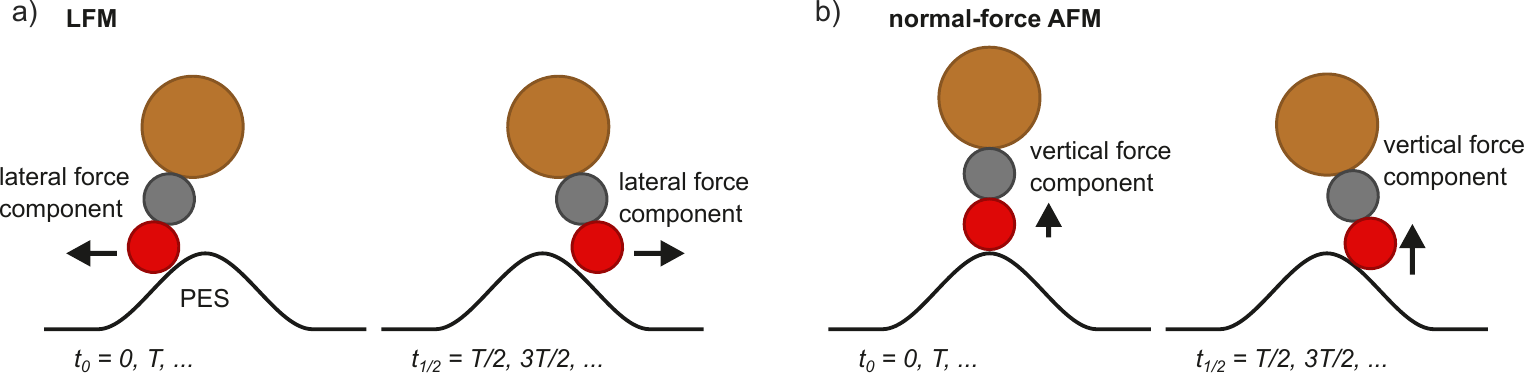}
	\caption{Schematic CO tip positions for a) LFM and b) normal-force AFM oscillating over a PES. The force components acting on the CO tip are indicated by black arrows for the oscillation extrema. }
	\label{FigS3} 
\end{figure}

To explain the enhanced sensitivity of \Df measured by LFM in comparison to \Df measured by AFM, one needs to understand the force components acting during oscillation. 
Figure \ref{FigS3} schematically displays the extreme positions of a CO tip during oscillation over a PES and the resulting force components (black arrows) for a) LFM and b) normal-force AFM. 
In the case of LFM, see Figure \ref{FigS3} a), the CO is located on the left-hand side of the PES for times $t_0 = (0,~T,~\dots)$ and its lateral force component points towards the left. For times $t_{1/2} = (T/2,~3T/2,~\dots)$, it is located at the right-hand side of the PES and its force component points towards the right. 
Within one oscillation, the CO therefore changes the side of the PES, i.e. snaps. 
The lateral force components have opposite direction during one oscillation, as indicated by black arrows, maximizing the resulting \Df: 

\begin{equation}
    \Delta f = - \frac{f_0^2}{k A} \int_0^T F (t) \cos(2\pi f_0 t) \text{d} t.
\end{equation}

where $F$ is the relevant force component (in the case of LFM, the force component in the $x$ direction), and $T=1/f_0$ is the period of oscillation. 
$k$ and $A$ have been previously defined.

In the case of normal-force AFM, see Figure \ref{FigS3} b), the CO tip is oscillating vertically directly over the maximum of the PES. 
At times $t_0 = (0,~T,~\dots)$, the tip reaches its maximal distance from the PES during one oscillation with the vertical force component pointing upwards. 
At times $t_{1/2} = (T/2,~3T/2,~\dots)$, the CO gets deflected on one side of the PES due to spatial constraint, but the vertical force component still points upwards. 

We propose that this is the reason for the enhanced sensitivity of the LFM signal and its ability to identify the atomic species on the copper oxide surface where normal-force AFM with a CO-tip has previously failed. 

\clearpage

\section{Frequency Shift Hysteresis}
\subsection{Experiment}

\begin{figure}
	\centering \includegraphics[width=1\linewidth]{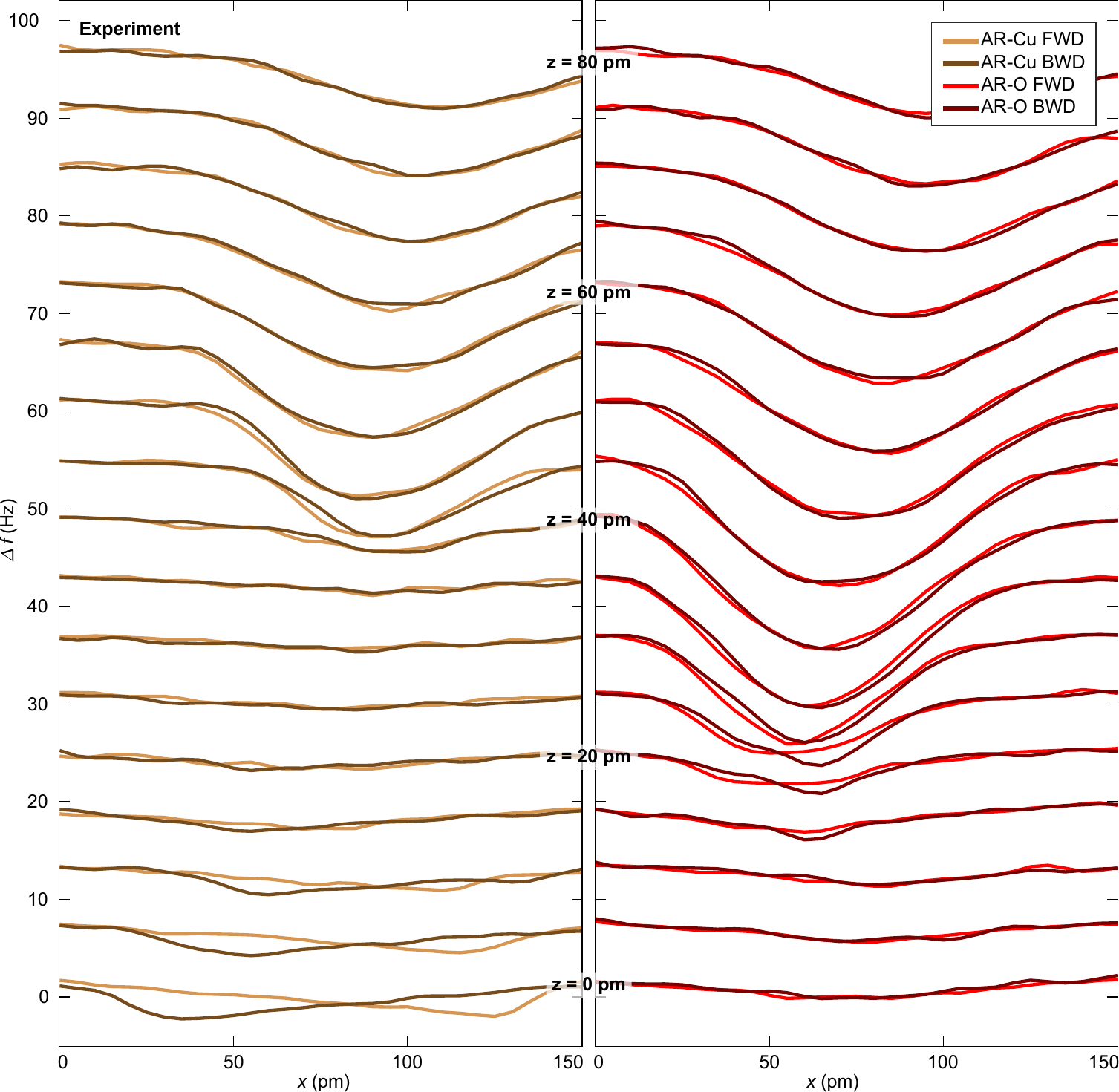}
	\caption{Experimental \Dfx forward and backward line scans at AR-Cu site (left) and AR-O site (right) for multiple $z$-values, as indicated. The $y$-location of the AR-Cu \Dfx slices are identical to that of figure 5b) in the main text. The individual slices are offset by a constant to increase readability. }
	\label{FigS2a} 
\end{figure}

To investigate the $z$-progression of the hysteresis, discussed in Figure 5 of the main text, \Dfx line scans are plotted in Figure \ref{FigS2a} for multiple values of $z$. 
The $z$-value is increased by $5$~pm for the individual curves and they are offset by a constant to increase readability. 
The AR-Cu traces for $z = 0$ are identical to the ones shown in Figure 5 b) of the main text. 
Focusing on the AR-Cu site first (left), it is observed that the hysteresis vanishes after approximately $15$~pm in $z$. 
At a height of approximately $40$~pm, a dip starts to appear.

For the AR-O site (right) no hysteresis is observed for any $z$-slice:
The curves overlap within the measurement range. 
The dip starts to be visible at lower heights, approximately $z=20$~pm. 

These observations are in agreement with those presented in Figure 2 in the main text. 
They highlights the short $z$-range of the measurement of the stick-slip mechanism. 

\subsection{Lennard-Jones Simulation}

\begin{figure}
	\centering \includegraphics[width=1\linewidth]{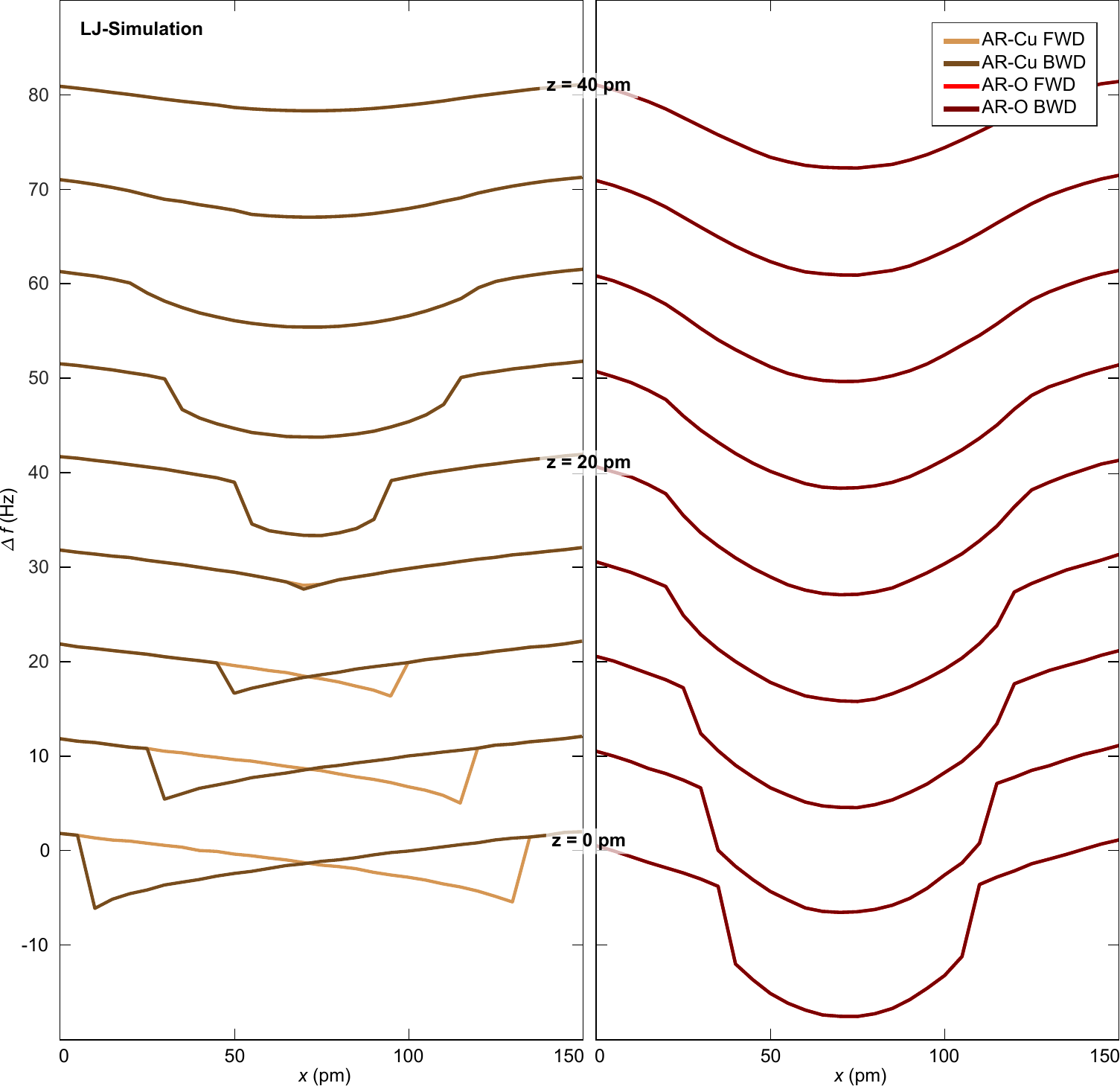}
	\caption{Lennard-Jones simulated \Dfx forward and backward line scans at AR-Cu site (left) and AR-O site (right) for multiple $z$-values, as indicated. The $y$-location of the AR-Cu \Dfx slices are identical to that of Figure 5c) in the main text. The individual slices are offset by a constant to increase readability.}
	\label{FigS2b} 
\end{figure}

Figure \ref{FigS2b} shows the corresponding L-J simulated \Dfx line scans for multiple $z$, as indicated. 
The $z$-value is increased by $5$~pm for the individual curves and they are offset by a constant to increase readability. 
The AR-Cu traces for $z=0$~pm are identical to the ones shown in Figure 5 c) of the main text. 
Similar to the experimental curves in Figure \ref{FigS2a}, the hysteresis seen at the AR-Cu site (left) vanishes after approximately $15$~pm. 
The prominent dip starts to appear at $z\approx20$~pm. 

\clearpage

\bibliography{CuOx}